\pdfoutput=1
\documentclass[
reprint,
groupedaddress,
showpacs,
bibnotes,
amsmath,amssymb,
prb,
floatfix,
]{revtex4-1}

\usepackage{graphicx}% Include figure files
\usepackage{dcolumn}% Align table columns on decimal point
\usepackage{bm}% bold math
\usepackage{setspace}

\usepackage{color}

\begin{document}
\newcommand{\BFA}{BaFe$_{2}$As$_{2}$}
\newcommand{\BKFA}{(Ba$_{1-x}$K$_x$)Fe$_{2}$As$_{2}$}
\newcommand{\CFPPA}{Ca$_{10}$(Fe$_{1-x}$Pt$_x$As)$_{10}($Pt$_{3}$As$_8$)}
\newcommand{\CFPA}{Ca$_{10}$(FeAs)$_{10}($Pt$_3$As$_8$)}
\newcommand{\CFPvA}{Ca$_{10}$(FeAs)$_{10}($Pt$_4$As$_8$)}

%\preprint{APS/123-QED}

\title{Structural and magnetic phase transitions in triclinic Ca$_{10}$(FeAs)$_{10}($Pt$_3$As$_8$)}

\author{T. St\"urzer$^1$}
\author{G. M. Friederichs$^1$}
\author{H. Luetkens$^2$}
\author{A. Amato$^2$}
\author{H.-H. Klauss$^2$}
\author{F. Winter$^3$}
\author{R. P\"ottgen$^3$}
\author{D.Johrendt$^1$}
\email{johrendt@lmu.de}
\affiliation{
{$^1$}{Department Chemie, Ludwig-Maximilians-Universit\"{a}t M\"{u}nchen, D-81377 M\"{u}nchen, Germany}\\
{$^2$}{Labor f\"{u}r Myonenspinspektroskopie, Paul Scherrer
Institute, CH-5232 Villigen PSI, Switzerland} \\
{$^3$}{Institut f\"ur Anorganische und Analytische Chemie, Universit\"at M\"unster, 48149 M\"unster, Germany}
}

\date{\today}%

\begin{abstract}

We report the structural and magnetic phase transition of triclinic {\CFPA}. High-resolution X-ray diffraction reveals splitting of the in-plane ($a,b$) lattice parameters at $T_s \approx $ 120 K. Platinum-doping weakens the distortion and shifts the transition temperature to 80 K in Ca$_{10}$(Fe$_{1-x}$Pt$_x$As)$_{10}($Pt$_{3}$As$_8$) with $x$ = 0.03. $\mu$SR experiments show the onset of magnetic order near $T_s$ and a broad magnetic phase transition. No symmetry breaking is associated to the structural transition in {\CFPA} in contrast to the other parent compounds of iron arsenide superconductors.

\end{abstract}


\maketitle

%Introduction

\begin{figure}[h]
\center{
\includegraphics[width=80mm]{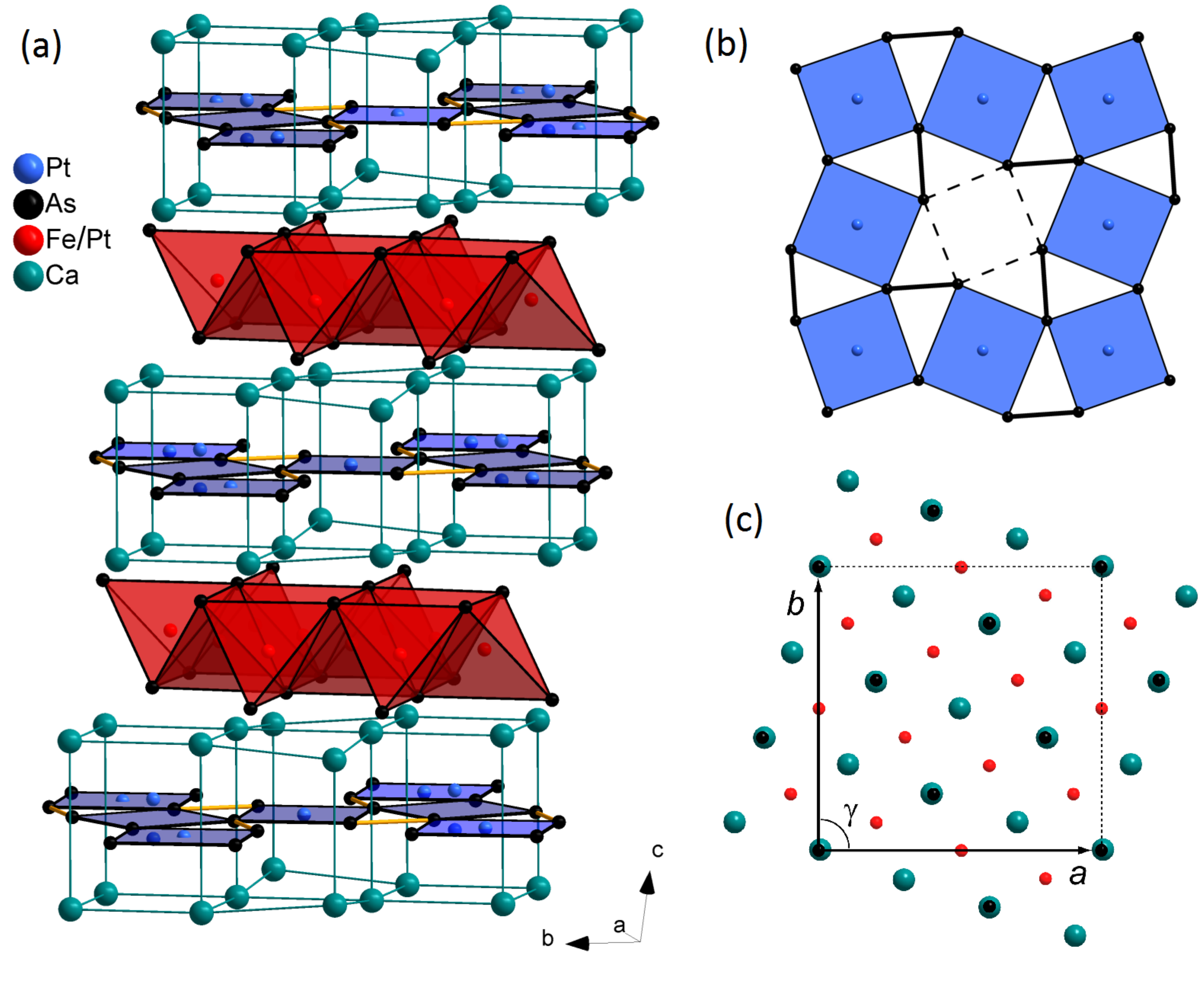}
\caption{(a) Crystal structure of {\CFPPA} ($x$ = 0, 0.03), (b) Pt$_3$As$_8$ layer, (c) FeAs layer with the in-plane lattice translations $a,b$.}
\label{fig:structure}
}
\end{figure}

Superconductivity in iron arsenide compounds emerges from stoichiometric parent compounds in the course of the destabilization of the antiferromagnetic ground states by chemical doping or pressure. The stripe-type antiferromagnetic ordering of the 1111-, 122-, and 111-type iron-arsenides is linked to an orthorhombic distortion of the tetragonal lattice,\cite{Cruz-2008,Rotter-2008,Tegel-2008,Parker-2009} which occurs at the temperature $T_s$ slightly above the Ne\'el-point $T_N$. This proximity of superconductivity to the structural and magnetic phase transition was not clearly evidenced in the more complex iron-arsenide  {\CFPPA}.\cite{Loehnert-2011,Cava-PNAS2011,Nohara-2011} Their crystal structures contain alternating layers of iron-arsenide and platinum-arsenide, each separated by calcium atoms as shown in Fig.~\ref{fig:structure}.

The compound referred to as the 1038 phase contains Pt$_3$As$_8$-layers, while in the 1048 phase one more platinum atom is located in Pt$_4$As$_8$-layers. Superconductivity in 1038-/1048-compounds is controlled by Pt-doping at the iron-sites and by doping with excess electrons either from the Pt$_4$As$_8$-layer or from La-doping at the Ca-sites. High critical temperatures up to 38 K only occur with clean FeAs-layers, while $T_c$ with Pt-doped layers remains below 15 K.\cite{Sturzer-2012} This doping behavior is similar to the other FeAs systems, where transition metal doping induces significantly lower critical temperatures than electron- or hole-doping outside the FeAs layers. Electronic structure calculations \cite{Loehnert-2011} as well as angle-resolved photoemission experiments \cite{Neupane-2011} have shown that the Fermi-surface of the 1038/1048-superconductors exhibits features very similar to the simpler FeAs compounds.

Thus far there is every indication that the 1038/1048-materials act up to the same principle than known FeAs compounds. Therefore a non-superconducting parent compound with antiferromagnetic ordering and  structural phase transition is expected to exist. We have recently proposed the stoichiometric 1038 phase {\CFPA} as the parent compound.\cite{Sturzer-2012} By assigning the usual ionic charges according to Ca$^{2+}_{10}$[(FeAs)$_{10}$]$^{10-}$(Pt$_3$As$_8$)$^{10-}$ we obtain the identical charge of the FeAs layer ($-1$) as in the 1111- or 122-type parent compounds. Moreover we found that superconductivity is induced from this stoichiometric 1038 phase by La-doping at the Ca-site. Thus one also expects an antiferromagnetic ground state of non-superconducting {\CFPA} and a structural distortion of the FeAs layer. The latter was also suggested on the base of polarized light imaging.\cite{Cho-2012} A recent preprint \cite{Ni-2012} reported the phase diagram of the La-doped 1038 phase and assumed the existence of structural magnetic transitions from weak features in the magnetic susceptibility, specific heat and kinks in the derivative of the electrical resistivity. Another preprint \cite{Zhou-2012} reported evidence for antiferromagnetic ordering in the 1038 phase from  $^{75}$As NMR data. All reports so far generally support the existence of a structural transition in the 1038 compound, but none of them gives clear experimental evidence of a lattice distortion.

In this letter we show that the non-superconducting 1038 phase undergoes a structural phase transition near 120 K. The splitting of the equal lattice parameters $a,b$ in the triclinic crystal structure is observed by high-resolution X-ray diffraction. Concomitant magnetic ordering is proved by means of $\mu$SR data showing an onset of magnetic ordering near $T_s$ followed by a broad magnetic transition.

Polycrystalline samples of platinum iron-arsenides were synthesized as described in Ref.\cite{Sturzer-2012}, and characterized by X-ray powder diffraction using the Rietveld method with TOPAS.\cite{Topas} Compositions were determined within errors of $\pm$10\% by refining occupation parameters and by X-ray spectroscopy (EDX). Temperature dependent X-ray powder diffraction data were collected using a HUBER G670 Guinier imaging plate diffractometer (Cu-K$\alpha_1$ radiation, Ge-111 monochromator) equipped with a close-cycle He-cryostat. Dc-resistivity was measured on a cold pressed pellet which has been annealed at 1073 K for 20 h. Magnetic susceptibility was measured using a Quantum Design MPMS-XL5 SQUID magnetometer. $\mu$SR measurements have been performed using the GPS and Dolly spectrometers located at the $\pi$M3 and $\pi$E1 beamlines of the Swiss Muon Source at the Paul Scherrer Institut, Switzerland. The data have been analyzed using the MUSRFIT package.\cite{Suter-2012}

\begin{figure}[h]
\center{
\includegraphics[width=80mm]{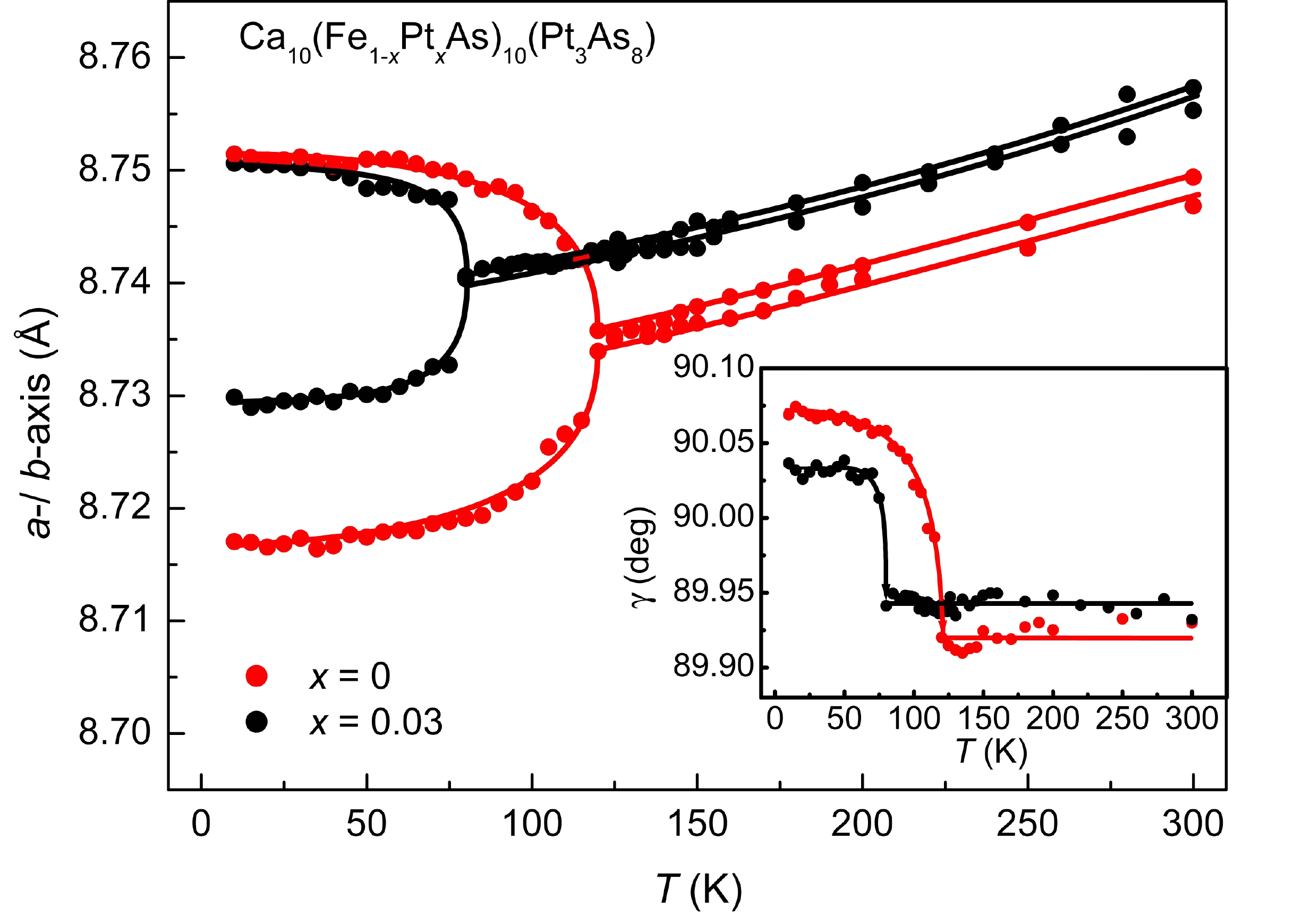}
\caption{Lattice parameters of {\CFPPA} ($x$ = 0, 0.03); The insert shows the angle $\gamma$ between $a,b$}
\label{fig:distortion}
}
\end{figure}

Fig.~\ref{fig:distortion} shows the temperature dependency of the lattice parameters refined from X-ray powder data for {\CFPPA} with $x$ = 0 and 0.03. Within the accuracy of the method, the lattice parameters $a$ and $b$ are equal at ambient temperature due to the square base plane of the structure. The $c$-axis decreases monotonically with cooling, while the angles remain nearly constant. The stoichiometric and underdoped 1038 compounds exhibit lattice distortions at 120 K and 80 K, respectively. Similar properties are known for a variety of iron arsenide compounds like BaFe$_2$As$_2$, where the phase transition results in a symmetry reduction from tetragonal $I4/mmm$ to orthorhombic $Fmmm$, or LaOFeAs with a transition from $P4/nmm$ to $Cmme$. Since the triclinic structure of the 1038 compounds precludes further symmetry reduction, the existence of the distortion indicates the loss of local tetragonal symmetry in the FeAs-layer being the essential effect of the structural phase transition.

\begin{figure}[h]
\center{
\includegraphics[width=90mm]{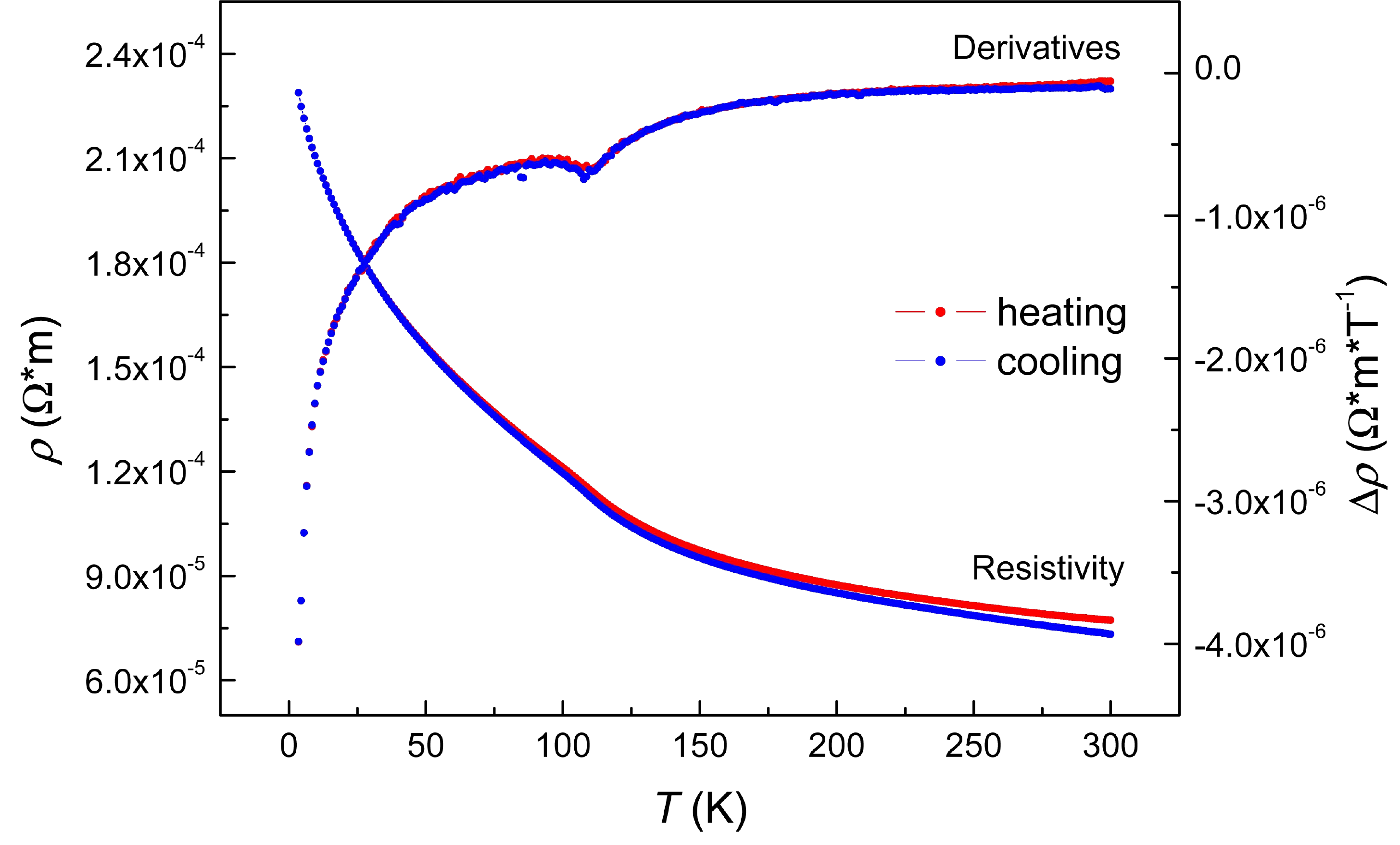}
\caption{DC resistivity of {\CFPA}; The derivative emphasizes the anomaly at 120 K.}
\label{fig:resistivity}
}
\end{figure}

The resistivity measurement is displayed in Fig.~\ref{fig:resistivity}. The semiconductor like development to low temperatures is in contradiction to known iron arsenides, but was recently also observed by Xiang \textit{et al.}.\cite{Xiang-2012} Absolute resistivity values are in the typical range of iron pnictides. The derivatives of resistivity data by temperature reveal an anomaly near 120 K coinciding for heating and cooling measurement.

\begin{figure}[h]
\center{
\includegraphics[width=80mm]{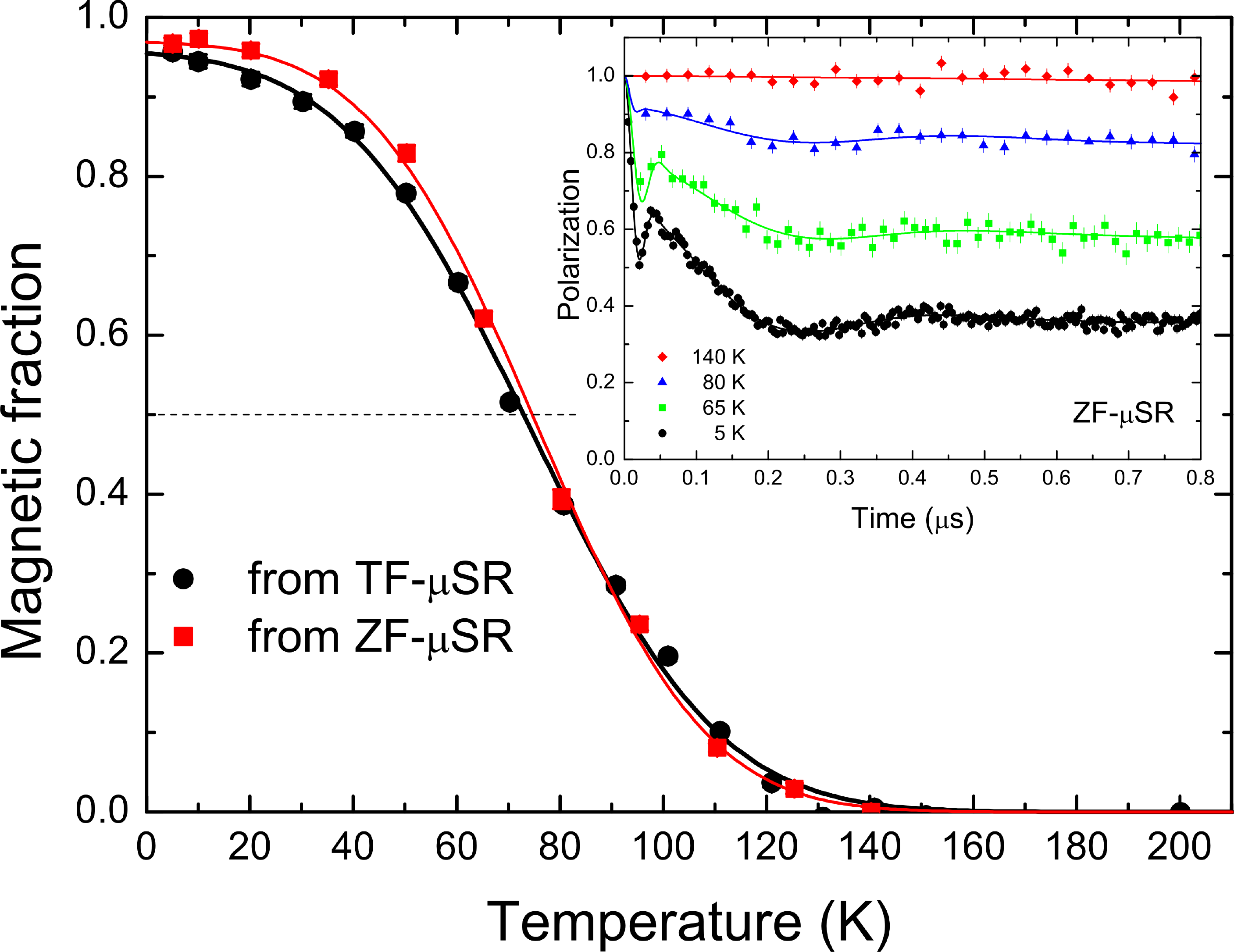}
\caption{Magnetic phase fraction of {\CFPA} obtained from TF-and ZF-$\mu$SR data; Insert: ZF-$\mu$SR spectra of {\CFPA} at different temperatures}
\label{fig:muSR}
}
\end{figure}

Muon spin rotation experiments with stoichiometric {\CFPA} detected three different muon precession frequencies with constant ratios, which accounts for three different muon sites in the magnetic unit cell of the homogenous phase. The onset of long range magnetic order below 130 K was found in TF and ZF modes. Thereby the magnetic order develops gradually, reaching 100\% at 5 K as shown in Fig.~\ref{fig:muSR}. The sample is 100\% and static magnetically ordered at 5 K, which is evident from the so-called 1/3 tail of the spectra which is not damped (insert in Fig.~\ref{fig:muSR}). Interestingly the $\mu$SR frequency is more or less temperature independent, which may indicate a first order phase transition.

\begin{figure}[h]
\center{
\includegraphics[width=80mm]{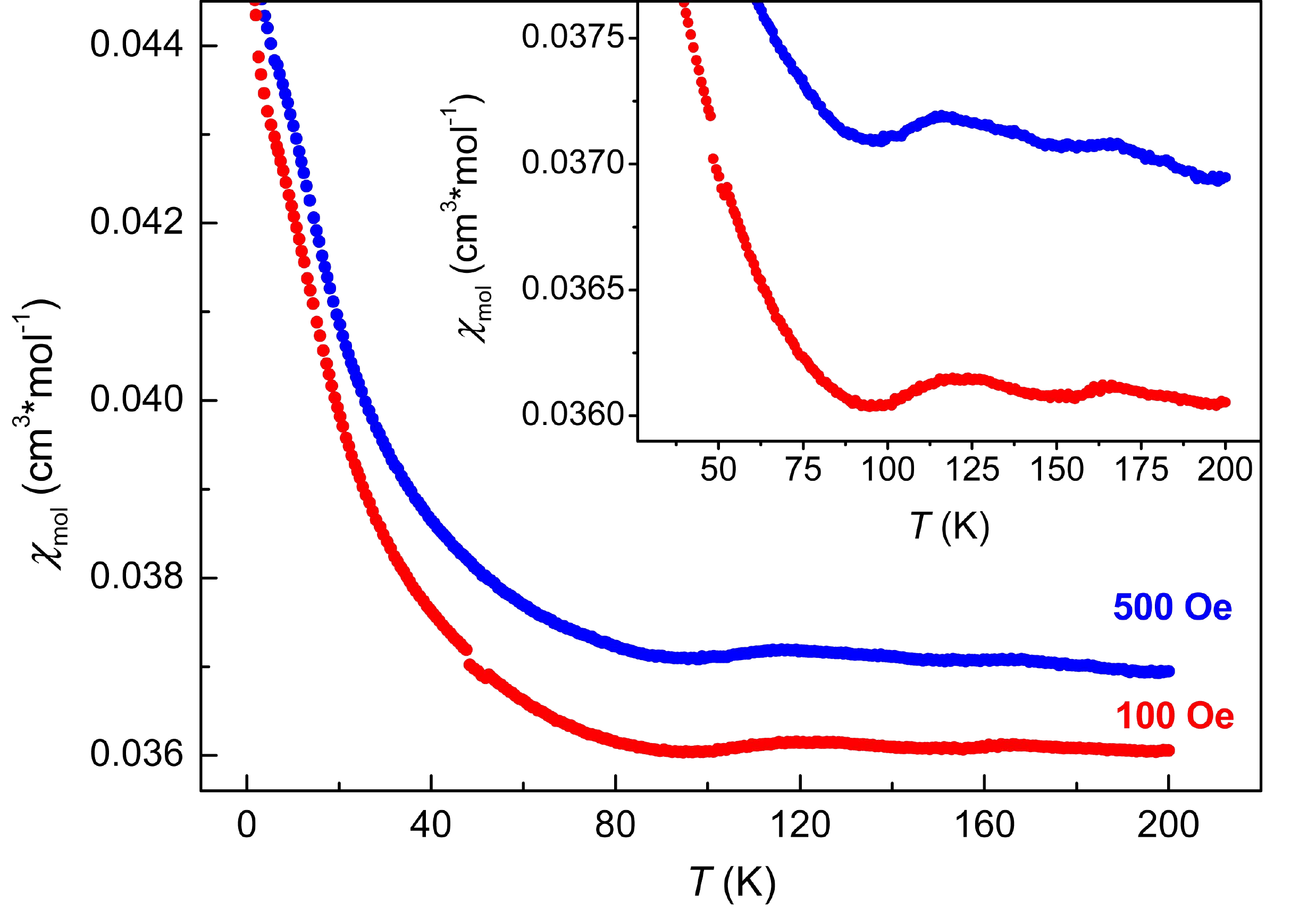}
\caption{Magnetic susceptibilty of {\CFPA}}
\label{fig:chi}
}
\end{figure}

The magnetic susceptibility of {\CFPA} (Fig.~\ref{fig:chi}) shows a weak and broad anomaly in the  temperature region of the structural transition in agreement with Ref.\cite{Ni-2012}, thus substantiating a gradual development of magnetic order. The linear magnetization isotherm at 1.8 K is compatible with antiferromagnetic order. From the results so far we suggest a magnetic ordered state similar to the parent compounds BaFe$_2$As$_2$ or LaOFeAs. However, the low space group symmetry allows deviations from the stripe-type pattern which remains to be seen.\\
~\\

~\\
In summary, our results give clear evidence for the magnetic and structural phase transition to an antiferromagnetic low temperature phase at $T_s \approx$ 120 K. In contrast to the 1111- and 122-type iron arsenides, the phase transition involves no reduction of the space group symmetry. Nevertheless the local tetragonal symmetry in the FeAs layers of {\CFPA} is broken. The magnitude of the lattice distortion is roughly half of that observed in BaFe$_2$As$_2$ in terms of the orthorhombic order parameter $\delta$ = ($a$-$b$)/($a$+$b$), and decreases with Pt-doping on the Fe-sites as expected. Finally the transition is completely suppressed in optimally doped La-1038. The onset of long range magnetic order in {\CFPA} coincides with the structural distortion at $T_s \approx$ 120 K. In contrast to BaFe$_2$As$_2$ the magnetic fraction develops gradually and reaches 100\% not until 5 K. Taken this together with the temperature independent $\mu$SR frequencies, a gradual increase of the structurally distorted compound at costs of the ambient temperature phase should be observable below 130 K. However, our low temperature structural data suggest a rather sharp structural change in the whole sample without coexistence with the undistorted phase. While the detailed nature of the phase transition necessitates further investigations, our results demonstrate that the 1038 material acts up to the same principle as the known FeAs superconductors with {\CFPA} as the parent compound.\\
~\\

\begin{acknowledgments}

This work was financially supported by the German Research Foundation (DFG) within SPP1458, grant No. JO257/6.

\end{acknowledgments}

\bibliographystyle{apsrev}

\bibliography{1038-SPT}

\end{document}